# MedCPT: Contrastive Pre-trained Transformers with Large-scale PubMed Search Logs for Zero-shot Biomedical Information Retrieval

Qiao Jin[1], Won Kim[1], Qingyu Chen[1], Donald C. Comeau[1], Lana Yeganova[1], W. John Wilbur[1], Zhiyong Lu[1]

[1]National Center for Biotechnology Information (NCBI), National Library of Medicine (NLM), National Institutes of Health (NIH)

Correspondence: zhiyong.lu@nih.gov

**Abstract**

Motivation

Information retrieval (IR) is essential in biomedical knowledge acquisition and clinical decision support. While recent progress has shown that language model encoders perform better semantic retrieval, training such models requires abundant query-article annotations that are difficult to obtain in biomedicine. As a result, most biomedical IR systems only conduct lexical matching. In response, we introduce MedCPT, a first-of-its-kind Contrastively Pre-trained Transformer model for zero-shot semantic IR in biomedicine.

Results

To train MedCPT, we collected an unprecedented scale of 255 million user click logs from PubMed. With such data, we use contrastive learning to train a pair of closely-integrated retriever and re-ranker. Experimental results show that MedCPT sets new state-of-the-art performance on six biomedical IR tasks, outperforming various baselines including much larger models such as GPT-3-sized cpt-text-XL. In addition, MedCPT also generates better biomedical article and sentence representations for semantic evaluations. As such, MedCPT can be readily applied to various real-world biomedical IR tasks.

Availability

The MedCPT code and API are available at https://github.com/ncbi/MedCPT.

## Introduction

Information retrieval (IR) is an important step in biomedical knowledge discovery and clinical decision support (Ely, et al., 2005; Gopalakrishnan, et al., 2019). However, most IR systems in biomedicine are keyword-based, which will miss articles that are semantically relevant but have no lexical overlap with the input query. Recent progress in IR and deep learning has shown that dense retrievers, which encode and match queries and documents as dense vectors, can perform better semantic retrieval than traditional sparse (lexical) retrievers such as BM25 (Karpukhin, et al., 2020; Khattab and Zaharia, 2020; Lin, et al., 2022; Nogueira and Cho, 2019). They are typically based on pre-trained transformers (Vaswani, et al., 2017), and are further fine-tuned with task-specific data. However, dense retrieval models trained on general datasets cannot generalize well to domain-specific IR tasks (Thakur, et al., 2021). Nevertheless, domain-specific datasets are limited in scale and diversity, restricting the creation of generalizable models (Roberts, et al., 2014; Roberts, et al., 2017; Tsatsaronis, et al., 2015; Voorhees, et al., 2021). As a result, there is a pressing need for pre-trained models that can perform well across various biomedical IR tasks.

In response, we propose bio<u>M</u>edical <u>C</u>ontrastive <u>P</u>re-trained <u>T</u>ransformers (MedCPT), a novel model trained with an unprecedented scale of 255M query-article pairs from PubMed search logs. MedCPT is the first biomedical IR model that includes a pair of retriever and re-ranker closely integrated by contrastive learning. Unlike previous separately developed models that have a discrepancy between the two modules (Gao, et al., 2021), MedCPT re-ranker is trained with the negative distribution sampled from the pre-trained MedCPT retriever. This matches the inference time article distribution where the MedCPT re-ranker is used to re-rank the articles returned by the MedCPT retriever. As shown in Figure 1, we perform zero-shot evaluation on a wide range of biomedical IR tasks. For document retrieval, MedCPT (330M) achieves state-of-the-art (SOTA) document retrieval performance on three individual biomedical tasks and the overall average in BEIR (Thakur, et al., 2021), outperforming much larger models such as Google's GTR-XXL (4.8B) (Ni, et al., 2021) and OpenAI's cpt-text-XL (175B) (Hirschman, et al., 2012). For article representation, we also show that the MedCPT article encoder sets new SOTA performance on the RELISH similar article dataset (Brown, et al., 2019) and the MeSH prediction task in SciDocs (Cohan, et al., 2020). For sentence representation, MedCPT performs the best or second best among compared methods on the BIOSESS (Sogancioglu, et al., 2017) and MedSTS (Wang, et al., 2020) for semantic evaluations. As such, MedCPT can be readily applied to a variety of biomedical applications such as searching relevant documents, retrieving similar sentences, recommending related articles, as well as providing domain-specific retrieval-augmentation for large language models (Jin, et al., 2023).

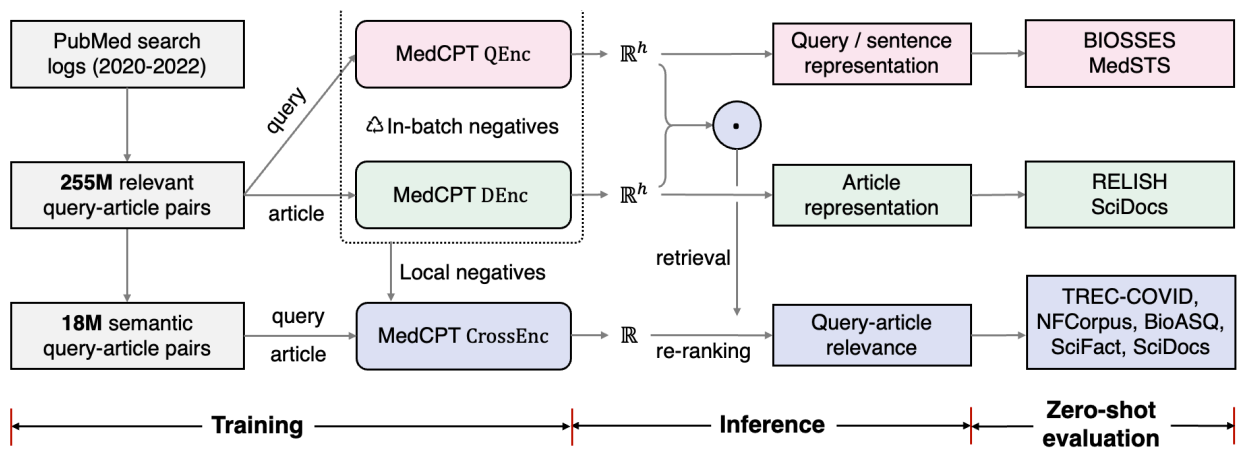

**Figure 1.** A high-level overview of this work. MedCPT contains a query encoder (QEnc), a document encoder (DEnc), and a cross-encoder (CrossEnc). The query encoder and document encoder compose of the MedCPT retriever, which is contrastively trained by 255M query-article pairs and in-batch negatives from PubMed logs. The cross-encoder is the MedCPT re-ranker, and is contrastively trained by 18M non-keyword query-article pairs and local negatives retrieved from the MedCPT retriever. MedCPT achieves state-of-the-art performance on various biomedical information retrieval tasks under zero-shot settings, including query-article retrieval, sentence representation, and article representation.

## Materials and Methods

### Query-article relevance data collection from PubMed search logs

We collected anonymous query-article clicks in PubMed search logs in three years (2020-2022) to train MedCPT. The raw logs contain 167M unique queries and 23M unique PubMed articles. We first filtered the navigational queries like author and journal title searches with Field Sensor (Yeganova, et al., 2018). After filtering, there are 87M informational queries and 17M articles. Based on the user click information, we generated 255M relevant query-article pairs to train the MedCPT retriever. However, most of such queries are short keywords, and matching them to the clicked articles is a relatively simple task. As such, we use a difficult subset that requires better semantic understanding to train the MedCPT re-ranker, which is aimed to distinguish harder negatives among the top-ranking articles returned by the retriever. Specifically, we further filtered out 79M keyword queries from the informational query set, which are defined as either having only one word or all of the clicked articles containing exact mentions of the whole input query. In the end, there are 7.7M non-keyword (e.g., short sentences) queries and 5.2M articles, from which we generated 18.3M relevant query-article pairs to train the MedCPT re-ranker.

### MedCPT architecture

MedCPT includes a first-stage retriever and a second-stage re-ranker. The retriever includes a query encoder (QEnc in Figure 1) and a document encoder (DEnc). This bi-encoder architecture is scalable because millions of articles can be encoded offline, and only one encoding computation for the query and a nearest neighbor search are required during real-time inference. The re-ranker is a cross-encoder (CrossEnc) that is computationally more expensive but also more accurate due to the cross-attention computation between query and article tokens. It will only be applied on the top articles returned by the retriever and generate the final article ranking.

### MedCPT retriever

The MedCPT retriever contains QEnc and DEnc, both of which are Transformer (Trm) encoders initialized by PubMedBERT (Gu, et al., 2021). It represents the query $q$ and document $d$ by:

$$E(q) = \text{QEnc}(q) = \text{Trm}([\text{CLS}]\, q\, [\text{SEP}])_{[\text{CLS}]} \in \mathbb{R}^h$$

$$E(d) = \text{DEnc}(d) = \text{Trm}([\text{CLS}]\, d^{\text{title}}\, [\text{SEP}]\, d^{\text{abstract}}\, [\text{SEP}])_{[\text{CLS}]} \in \mathbb{R}^h$$

where [CLS] and [SEP] are the special tokens used in BERT. $h$ is the hidden. $d^{\text{title}}$ and $d^{\text{abstract}}$ denote the title and abstract. Then, the relevance is calculated as: $Rel(q, d) = E(q)^T E(d) \in \mathbb{R}$.

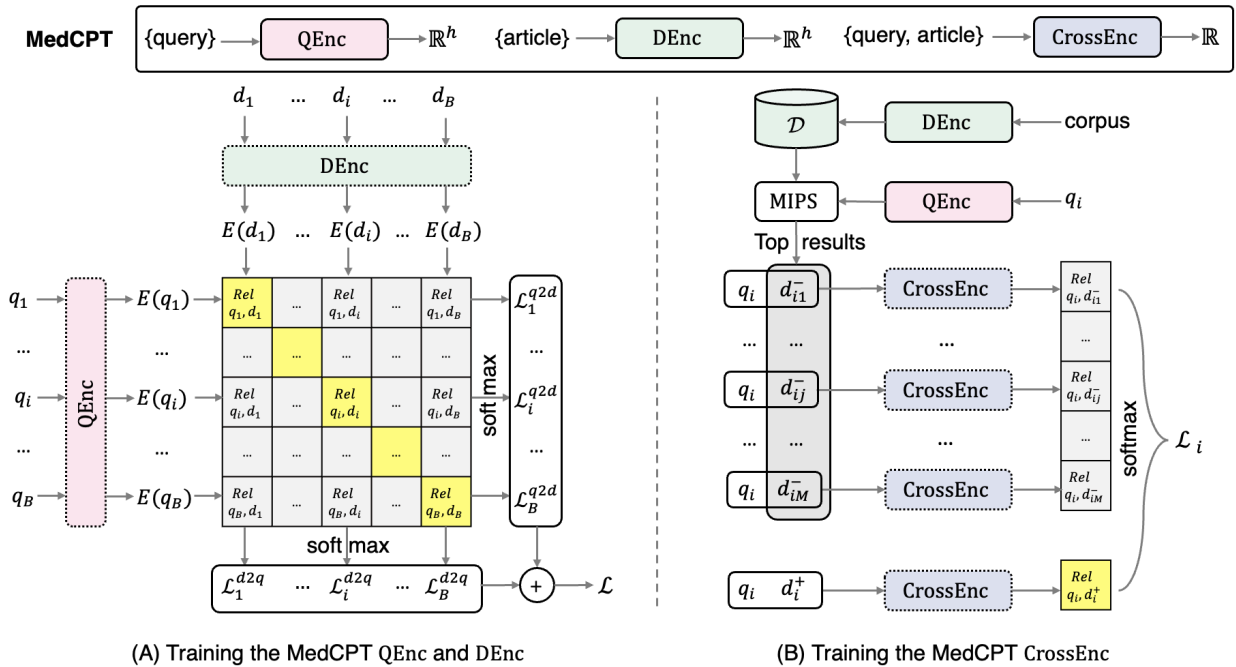

(A) Training the MedCPT QEnc and DEnc  (B) Training the MedCPT CrossEnc

**Figure 2.** Overview of the MedCPT training process. (A) Training the MedCPT query encoder (QEnc) and document encoder (DEnc) using a contrastive loss with query-document pairs and in-batch negatives; (B) Training the MedCPT cross-encoder

(CrossEnc) using a contrastive loss with non-keyword query-article pairs and local negatives derived from the MedCPT retriever. Models in dashed and solid lines denote un-trained and pre-trained, respectively. MIPS: maximum inner product search.

As shown in Figure 2 (A), to train the MedCPT retriever, each instance has a query $q$, a clicked document $d$, and the number of clicks $c$. Each mini-batch contains $|B|$ instances, denoted as $[q_i, d_i, c_i]_{i=1}^{|B|}$. We use contrastive loss with in-batch negatives (Karpukhin, et al., 2020; Neelakantan, et al., 2022). Specifically, we first generate all $E(q_i)$ and $E(d_i)$ from $\text{QEnc}$ and $\text{DEnc}$, where $d_i$ is a relevant document for $q_i$. We assume that the $(|B| - 1)$ other documents $[d_j \mid j \neq i]$ in the mini-batch are irrelevant documents for $q_i$. Similarly, we also consider the $(|B| - 1)$ other queries $[q_j \mid j \neq i]$ in the mini-batch as irrelevant queries for $d_i$. For the training instance $i$, we calculate its query-to-document loss $\mathcal{L}_i^{q2d}$ and document-to-query loss $\mathcal{L}_i^{d2q}$ by:

$$\mathcal{L}_i^{q2d} = -\log \left( \frac{\exp(E(q_i)^T E(d_i))}{\sum_{m=1}^{|B|} \exp(E(q_i)^T E(d_m))} \right) \text{ and } \mathcal{L}_i^{d2q} = -\log \left( \frac{\exp(E(q_i)^T E(d_i))}{\sum_{m=1}^{|B|} \exp(E(q_m)^T E(d_i))} \right)$$

We further weight the loss of instances by their clicks: $\mathcal{L}_B^{q2d} = \sum_{i=1}^{|B|} w_i \mathcal{L}_i^{q2d}$ and $\mathcal{L}_B^{d2q} = \sum_{i=1}^{|B|} w_i \mathcal{L}_i^{d2q}$, where $w_i = \frac{\log_2(c_i+1)}{\sum_{k=1}^{|B|} \log_2(c_k+1)}$. The final loss $\mathcal{L}_B$ of the mini-batch is their weighted sum and is optimized by gradient-based methods.

### MedCPT re-ranker

The MedCPT re-ranker is a cross-encoder, denoted as $\text{CrossEnc}$. Similarly, $\text{CrossEnc}$ is also initialized with PubMedBERT. The MedCPT re-ranker predicts the relevance between a query $q$ and a document $d$ by passing them into a single $\text{CrossEnc}$. Specifically,

$$Rel(q, d) = \text{CrossEnc}(q, d) = W^T \text{Trm}([\text{CLS}]\ q\ [\text{SEP}]\ d\ [\text{SEP}])_{[\text{CLS}]} + b \in \mathbb{R}$$

where $W \in \mathbb{R}^h$ and $b \in \mathbb{R}$ are trainable parameters.

As shown in Figure 2 (B), for training the MedCPT re-ranker, each instance has a query $q_i$, a clicked document $d_i^+$, and a list of $M$ irrelevant (not clicked) documents $\{d_{ij}^- \mid j = 1, 2, 3, \dots, M\}$. Following (Gao, et al., 2021), we use local negatives to train the MedCPT re-ranker instead of in-batch negatives. Specifically, unlike the in-batch negative documents used by the MedCPT retriever that are approximately random samples, the local negative documents are sampled from rank $e$ to rank $f$ in the top retrieved documents by the pre-trained MedCPT retriever through a maximum inner product search, which ensures that the MedCPT re-ranker can distinguish the hard negatives returned by the retriever. The loss $\mathcal{L}_i$ for the instance is a negative log-likelihood loss:

$$\mathcal{L}_i = -\log \left( \frac{\exp\left(\text{CrossEnc}(q_i, d_i^+)\right)}{\exp\left(\text{CrossEnc}(q_i, d_i^+)\right) + \sum_{j=1}^{M} \exp\left(\text{CrossEnc}(q_i, d_{ij}^-)\right)} \right)$$

We take a weighted sum of the instance-level loss and optimize the final loss by gradient-based methods. More details on MedCPT inference and configuration are shown in Appendix A.

## Results

### MedCPT achieves state-of-the-art performance on biomedical IR tasks

Benchmarking-IR (BEIR) (Thakur, et al., 2021) is a standardized evaluation benchmark for zero-shot IR systems. We evaluate MedCPT with all five biomedical tasks in the BEIR benchmark. Appendix C describes the evaluation details and Table 1 shows the evaluation results.

First, MedCPT improves its initialization PubMedBERT by huge margins, where the latter basically fails on the retrieval tasks. Overall, MedCPT sets new SOTA performance on 3/5 tasks, surpassing compared sparse (Dai and Callan, 2020; Nogueira, et al., 2019; Zhang, et al., 2015) , dense (Hofstätter, et al., 2021; Izacard, et al., 2021; Karpukhin, et al., 2020; Xiong, et al., 2020), and late-interaction (Khattab and Zaharia, 2020) retrievers on all of the compared tasks. As shown in the BEIR paper, BM25 is a strong baseline that is generalizable to biomedical IR tasks. Notably, MedCPT is still better than BM25 with cross-encoder in 4/5 of the evaluated tasks, showing its effectiveness at retrieving relevant articles for biomedical queries. BM25 with re-ranker is only better on the TREC-COVID dataset, which might be due to annotation biases (Thakur, et al., 2021). We further compare MedCPT with more recent large dual retriever models, represented by Google's GTR and OpenAI's cpt-text, both of which have model sizes ranging from millions to billions of parameters. MedCPT is able to outperform all sizes of the GTR model. While the GPT-3 (Brown, et al., 2020) sized (175B) cpt-text-XL is better than MedCPT on NFCorpus, MedCPT outperforms cpt-text-XL on TREC-COVID and SciFact despite being about 500 times smaller. This indicates that small models trained on domain-specific datasets can still have better in-domain zero-shot performance than much larger general domain retrievers.

**Table 1.** Zero-shot performance of MedCPT on biomedical subtasks of the BEIR benchmark. **Bolded numbers**, underlined, and *italicized* numbers denote the **highest**, 2nd highest, and *3rd highest*, respectively. COVID: TREC-COVID; NFC: NFCorpus; Avg.: average.

| Method | Size | COVID | NFC | BioASQ | SciFact | SciDocs | Avg. |
|---|---|---|---|---|---|---|---|
| **Sparse retrievers** | | | | | | | |
| BM25 | - | 0.656 | 0.325 | 0.465 | 0.665 | 0.158 | *0.454* |
| BM25 + MiniLM | 66M | **0.757** | 0.350 | <u>0.523</u> | 0.688 | *0.166* | <u>0.497</u> |
| DeepCT | 110M | 0.406 | 0.283 | 0.407 | 0.630 | 0.124 | 0.370 |
| SPARTA | 110M | 0.538 | 0.301 | 0.351 | 0.582 | 0.126 | 0.380 |
| docT5query | 220M | 0.713 | 0.328 | 0.431 | 0.675 | 0.162 | 0.462 |
| **Dense retrievers** | | | | | | | |
| DPR | 110M | 0.332 | 0.189 | 0.127 | 0.318 | 0.077 | 0.209 |
| ANCE | 110M | 0.654 | 0.237 | 0.306 | 0.507 | 0.122 | 0.365 |
| TAS-B | 66M | 0.481 | 0.319 | 0.383 | 0.643 | 0.149 | 0.395 |
| GenQ | 220M | 0.619 | 0.319 | 0.398 | 0.644 | 0.143 | 0.425 |
| Contriever | 110M | 0.596 | 0.328 | - | 0.677 | 0.165 | - |
| Contriever + MiniLM | 176M | *0.701* | 0.344 | - | 0.692 | <u>0.171</u> | - |
| ColBERT | 110M | 0.677 | 0.305 | *0.474* | 0.671 | 0.145 | *0.454* |
| **Large language model retrievers** | | | | | | | |
| Google GTR-Base | 110M | 0.539 | 0.308 | 0.271 | 0.600 | 0.149 | 0.373 |
| Google GTR-Large | 335M | 0.557 | 0.329 | 0.320 | 0.639 | 0.158 | 0.401 |
| Google GTR-XL | 1.24B | 0.584 | 0.343 | 0.317 | 0.635 | 0.159 | 0.408 |
| Google GTR-XXL | *4.80B* | 0.501 | 0.342 | 0.324 | 0.662 | 0.161 | 0.398 |
| OpenAI cpt-text-S | 300M | 0.679 | 0.332 | - | 0.672 | - | - |
| OpenAI cpt-text-M | 1.20B | 0.585 | *0.367* | - | 0.704 | - | - |
| OpenAI cpt-text-L | <u>6.00B</u> | 0.562 | <u>0.380</u> | - | *0.744* | - | - |
| OpenAI cpt-text-XL | **175B** | 0.649 | **0.407** | - | <u>0.754</u> | - | - |
| **MedCPT** | | | | | | | |
| MedCPT | 330M | <u>0.709</u> | 0.355 | **0.553** | **0.761** | **0.172** | **0.510** |
| MedCPT (retriever only) | 220M | 0.697 | 0.340 | 0.332 | 0.724 | 0.123 | 0.443 |
| MedCPT w/o contrastive pre-training (PubMedBERT) | 110M | 0.059 | 0.015 | - | 0.010 | 0.004 | - |

MedCPT generates better biomedical article representations

We evaluate the MedCPT article encoder on the RELISH article similarity task (Brown, et al., 2019). RELISH is an expert-annotated dataset that contains 196k article-article relevance annotations for 3.2k query articles, as described in Appendix D. Table 2 shows the evaluation results on RELISH. The MedCPT article encoder ($DEnc$) outperforms all other models, including SPECTER (Cohan, et al., 2020) and SciNCL (Ostendorff, et al., 2022) that are specifically trained with article-article citation information. Compared to its base PubMedBERT model, the MedCPT article encoder improves by over 10% performance. We also evaluate the MedCPT article encoder on SciDocs (Cohan, et al., 2020) as described in Appendix E, which contains all scientific domains from biomedicine to engineering. The MedCPT article encoder achieves SOTA performance on the MeSH prediction subtask and is comparable to SOTA methods on the overall score, showing its effectiveness on biomedical tasks and generalizability to other scientific domains.

**Table 2.** Evaluation results of the MedCPT article encoder on the RELISH dataset. **Bolded numbers**, underlined, and *italicized* numbers denote the **highest**, 2nd highest, and *3rd highest*, respectively. All numbers are percentages. Avg.: average.

| Method | MAP | | | NDCG | | | Avg. |
|---|---|---|---|---|---|---|---|
| | @5 | @10 | @15 | @5 | @10 | @15 | |
| Random | 79.33 | 77.22 | 75.41 | 80.70 | 77.67 | 76.40 | 77.79 |
| **Sparse retrievers** | | | | | | | |
| BM25 | 88.91 | 86.72 | 84.54 | 89.48 | 87.39 | 86.21 | 87.21 |
| PMRA | 90.30 | 87.57 | 85.75 | 90.95 | 88.40 | 87.45 | 88.40 |
| **Non-BERT embedding-based models** | | | | | | | |
| fastText | 85.75 | 82.81 | 81.79 | 86.79 | 83.79 | 83.12 | 84.01 |
| BioWordVec | 89.84 | 86.51 | 84.67 | 89.90 | 86.67 | 85.53 | 87.19 |
| InferSent | 85.21 | 82.16 | 80.41 | 86.56 | 83.31 | 82.35 | 83.33 |
| WikiSentVec | 87.92 | 85.23 | 83.40 | 88.65 | 85.74 | 84.81 | 85.96 |
| BioSentVec | 90.76 | 88.10 | 86.16 | 90.05 | 87.76 | 86.89 | 88.29 |
| LDA | 85.44 | 82.66 | 80.36 | 86.51 | 82.91 | 81.31 | 83.20 |
| Doc2Vec | 86.23 | 84.74 | 83.39 | 86.55 | 84.70 | 84.09 | 84.95 |
| **BERT-based models** | | | | | | | |
| BioBERT | 88.14 | 85.81 | 83.90 | 88.97 | 86.29 | 85.10 | 86.37 |
| PubMedBERT | 83.69 | 81.07 | 79.53 | 85.47 | 82.39 | 81.41 | 82.26 |
| SPECTER | *92.27* | *90.00* | *88.36* | *91.47* | *89.12* | *88.42* | *89.94* |
| SciNCL | 94.72 | 92.74 | 91.14 | 93.67 | 91.91 | 90.94 | 92.52 |
| MedCPT DEnc | **95.58** | **93.99** | **92.39** | **94.78** | **93.12** | **92.43** | **93.72** |

### MedCPT generates better biomedical sentence representations

We evaluate the MedCPT query encoder on two datasets for sentence similarities: BIOSSES in the biomedical domain (Sogancioglu, et al., 2017) and MedSTS in the clinical domain (Wang, et al., 2020). Appendix F introduces the evaluation details and Table 3 shows the evaluation results. On BIOSSES, MedCPT performs the best among all compared models, surpassing the second SciNCL by 5% relative performance (0.893 vs. 0.847). On the MedSTS dataset, MedCPT ranks the second and the performance is comparable to the highest-ranking model BioSentVec (Chen, et al., 2019) (0.765 vs. 0.767), which uses an external clinical corpus MIMIC-III (Johnson, et al., 2016) for its model training. Overall, our results show that the MedCPT query encoder can effectively encode biomedical and clinical sentences that reflect their semantic similarities.

| Model | BIOSSES | MedSTS |
|---|---|---|
| **Non-BERT embedding-based models** | | |
| BioWordVec | 0.694 | 0.747 |
| USE | 0.345 | 0.714 |
| BioSentVec (PubMed) | *0.817* | 0.750 |
| BioSentVec (MIMIC-III) | 0.350 | *0.759* |
| BioSentVec (PubMed + MIMIC-III) | 0.795 | **0.767** |
| **BERT-based models** | | |
| PubMedBERT | 0.528 | 0.521 |
| Clinical BERT | 0.556 | 0.525 |
| SPECTER | 0.694 | 0.702 |
| SciNCL | 0.847 | 0.706 |
| MedCPT QEnc | **0.893** | 0.765 |

**Table 3.** Evaluation results (Pearson's correlation coefficients) of the MedCPT query encoder on the BIOSSES and MedSTS datasets. **Bolded numbers**, underlined, and *italicized* numbers denote the **highest**, 2nd highest, and *3rd highest*, respectively. All numbers are percentages.

### Discussions

MedCPT is only trained with query-article click data derived from PubMed user logs, but it generalizes well and achieves the SOTA performance on many biomedical IR tasks in the BEIR benchmark, which indicates that query-article pairs in the PubMed search logs can serve as high-quality training data for serving general-purpose information needs in biomedicine. Furthermore, while not being explicitly trained with query similarity and article similarity data, the MedCPT query encoder and article encoder still achieve the SOTA performance on sentence similarity and article similarity tasks, respectively. This shows that the contrastive objective can train not only a dense retriever, but can also train the individual query and document encoders to perform tasks related to information-seeking behaviors. As such, MedCPT has broad implications in a variety of real-world scenarios: enhancing algorithms for biomedical literature search such as PubMed's Best Match (Fiorini, et al., 2018), where case studies in Appendix G show that MedCPT retrieves more semantically relevant articles than other commonly used literature search engines; improving similar article recommendation algorithms in literature search (Lin and Wilbur, 2007); facilitating sentence-to-sentence retrieval tasks such as sentence-level literature search (Allot, et al., 2019).

Although transformer-based retrieval and re-ranking models such as MedCPT can return more comprehensive results, they are not as controllable or explainable as sparse retrievers such as BM25. For example, when user searches the gene "MAP3K3", MedCPT will also return articles that only contain "MAP3K7", which might not be the original information need. In addition, the semantic similarity scores between a query article pair are not explainable. As such, one potential future direction is to develop hybrid dense-sparse retrieval systems that can harvest the advantages from both approaches (Ma, et al., 2020; Shin, et al., 2023).

To summarize, we use large-scale PubMed logs to contrastively train MedCPT, the first integral retriever-reranker model for biomedical information retrieval. Systematic zero-shot evaluations show that MedCPT achieves the highest performance for six different biomedical information retrieval tasks, including query-to-article retrieval, semantic article and sentence representation. We anticipate that MedCPT will have a broad range of applications and significantly enhance access to biomedical information, making it a valuable tool for researchers and practitioners alike.

### Acknowledgments
This research was supported by the NIH Intramural Research Program, National Library of Medicine.

# Appendix A: MedCPT Inference

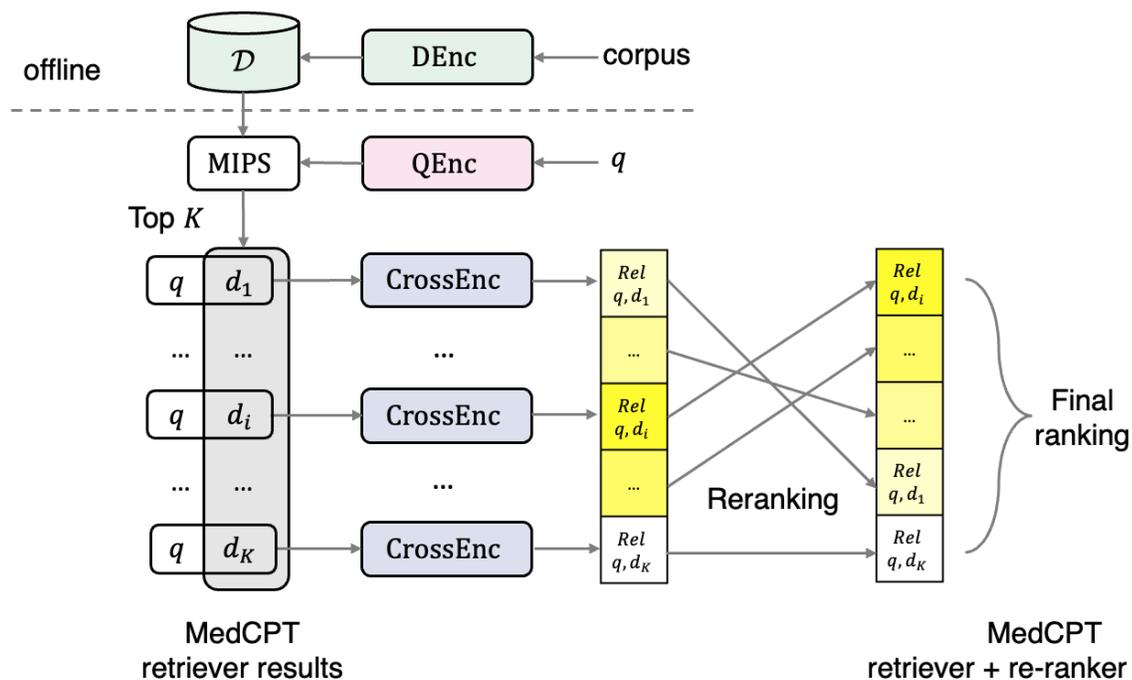

**Figure S1.** Architecture for inference with pre-trained MedCPT. A test query $q$ is fed to the MedCPT query encoder, and its representation will be matched against the corpus representations using MIPS. The corpus representations are calculated and saved off-line. Top $K$ retrieved articles will be re-ranked by the MedCPT cross-encoder. MIPS: maximum inner product search.

As shown in Figure S1, when applying the pre-trained MedCPT to downstream tasks in a zero-short fashion, we need to first encode the task corpus. Specifically, we use $\text{DEnc}$ to process each article $d_i$ in the task corpus, getting their representations $E(d_i) \in \mathbb{R}^h$. We save the representations of the entire corpus, denoted as:

$$\mathcal{D} = [E(d_1), E(d_2), \ldots, E(d_N)] \in \mathbb{R}^{N \times h}$$

where $N$ is the size of articles in the corpus. This step only needs to be done once in the offline setting.

Then, each input query $q$ is fed into $\text{QEnc}$, which generates its representative $E(q)$. We conduct a MIPS between $E(q)$ and $\mathcal{D}$ to find the top-$K$ most similar articles to the input query:

$$d_1^q, d_2^q, \ldots, d_K^q = \text{MIPS}(E(q), \mathcal{D})$$

We apply $\text{CrossEnc}$ to score the relevance between $q$ and each relevant article candidates in $d_1^q, d_2^q, \ldots, d_K^q$ retrieved from the previous step:

$$Rel(q, d_i^q) = \text{CrossEnc}(q, d_i^q)$$

Finally, we sort the retrieved articles by $Rel(q, d_i^q)$ from the highest to the lowest and return the sorted results.

We implemented the MedCPT using PyTorch (Paszke, et al., 2019) and the Hugging Face transformers library (Wolf, et al., 2020). The hidden dimension for MedCPT $h = 768$ as in the BERT-base configuration. We use the Adam optimizer (Kingma and Ba, 2014) without weight decay to train both the retriever and the re-ranker, where we set the learning rate 2e-5 and epsilon 1e-8. For the MedCPT retriever, $B = 32$ and $\alpha = 0.8$, and we also apply gradient accumulation of 8 steps. We train the retriever for 100k steps with 10k warm-up steps. For the MedCPT re-ranker: $M = 31$, $e = 50$, $f = 200$. We train the re-ranker for 10k steps with 1,000 warm-up steps. We apply cosine learning rate schedule after the warm-up steps. During inference, $N$ and $K$ vary for specific tasks. We implemented MIPS with the FlatIP index of the Faiss library (Johnson, et al., 2019).

## Appendix B: Compared methods

*Sparse retrievers*

Sparse retrievers, as known as lexical retrievers, match the queries and documents with overlapped terms. BM25 (Robertson and Zaragoza, 2009), a widely used lexical retriever, represents queries and documents as bag-of-words, and scores the relevance based on term frequency and inverse document frequency. DeepCT (Dai and Callan, 2020) uses contextualized term weights predicted by BERT. SPARTA (Zhang, et al., 2015) pre-computes contextualized matching weights for each possible term and the document, resulting in a sparse vector that has the size dimension as the BERT vocabulary. DocT5query (Nogueira, et al., 2019) performs document expansion with T5-generated (Raffel, et al., 2020) queries for the document. The PubMed Related Articles algorithm (Lin and Wilbur, 2007) uses a probabilistic topic-based model to compute the content similarity between two given articles.

*Dense retrievers*

Dense retrievers first encode both the queries and documents into low-dimensional dense vectors, and then perform nearest neighbor search to find the relevant documents for a given query. Depending on the encoder architecture, we broadly classify them into non-BERT embedding models and BERT-based dense retrievers.

Non-BERT embedding models include BioWordVec (Zhang, et al., 2019) which is a biomedical version of word2vec (Mikolov, et al., 2013), FastText (Bojanowski, et al., 2017), a linear model on the N-gram features of the input texts, Sent2vec (Pagliardini, et al., 2018) and its biomedical version BioSentVec (Chen, et al., 2019), the LDA topic model (Blei, et al., 2003), doc2vec (Le and Mikolov, 2014), InferSent (Conneau, et al., 2017), and Universal Sentence Encoder (USE) (Cer, et al., 2018). BERT-based dense retrievers: DPR (Karpukhin, et al., 2020) is a bi-encoder retriever trained by in-batch negatives and BM25 hard negatives. ANCE (Xiong, et al., 2020) improves the DPR training by using hard negatives from an approximate nearest neighbor index of the corpus. TAS-B (Hofstätter, et al., 2021) is a bi-encoder retriever distilled from a cross-encoder and ColBERT (Khattab and Zaharia, 2020) with balanced topic aware sampling. GenQ (Thakur, et al., 2021) is domain-adaptation method that trains a dense retriever with synthetic query-document pairs generated by T5 (Raffel, et al., 2020). Contriever (Izacard, et al., 2021) is a contrastively pre-trained dense retriever in the general domain with carefully engineered positive and negative query-document pairs. ColBERT (Khattab and Zaharia, 2020) is a late-interaction retriever that computes and matches the contextualized representations of each token in the query and document. In addition, we compare MedCPT with several off-the-shelf BERT models that are biomedical domain-specific, including its base model PubMedBERT (Gu, et al., 2021), BioBERT (Lee, et al., 2020), SPECTER (Cohan, et al., 2020), and SciNCL (Ostendorff, et al., 2022).

*Large language model retrievers*

We also compare MedCPT with two large language model retrievers: Google's GTR (Ni, et al., 2021) and OpenAI's cpt-text (Hirschman, et al., 2012). Unlike most dense retrievers that are based on the BERT-base model of 110M parameters, GTR and cpt-text use much larger language model encoders. Specifically, GTR is based on T5 (Raffel, et al., 2020) and its largest variant has 4.8B parameters, while cpt-text is based on GPT-3 (Brown, et al., 2020) and its largest variant has 175B parameter. Both GTR and cpt-text are pre-trained by large-scale Web corpora with in-batch negatives, and are further fine-tuned with supervised datasets such as MS MARCO (Bajaj, et al., 2016). In comparison, MedCPT is only trained by the user click data from PubMed logs without using any supervised datasets.

## Appendix C: Evaluation details on BEIR

We evaluate MedCPT on five biomedical tasks in the BEIR benchmark: TREC-COVID (Voorhees, et al., 2021), NFCorpus (Boteva, et al., 2016), BioASQ (Tsatsaronis, et al., 2015), SciFact (Wadden, et al., 2020), and SciDocs (Cohan, et al., 2020). TREC-COVID (Voorhees, et al., 2021) contains questions about the COVID-19 pandemic and uses the CORD-19 corpus (Wang, et al., 2020) as the document collection for retrieval. NFCorpus (Boteva, et al., 2016) collects natural language queries and relevant articles from the NutritionFacts.org site. BioASQ (Tsatsaronis, et al., 2015) is a community challenge for biomedical question answering, where the task used in BEIR is to retrieve relevant articles from PubMed for a given question. SciFact (Wadden, et al., 2020) is a scientific claim verification dataset, which contains a retrieval subtask and a veracity prediction subtask. BEIR uses the retrieval subtask of SciFact, where the objective is to find relevant articles that can be used to verify a given claim. SciDocs (Cohan, et al., 2020) is a benchmark for evaluating scientific article representation models. BEIR uses its citation prediction subtask, where the goal is to retrieve relevant citations for a given article. We use the official evaluation library for BEIR and report the normalized discounted cumulative gain at rank 10 (NDCG@10).

We compared it with various baselines, including its initialization model PubMedBERT (Gu, et al., 2021), BM25 (Robertson and Zaragoza, 2009) and BM25 with the MiniLM re-ranker (Wang, et al., 2020), sparse retrievers such as DeepCT (Dai and Callan, 2020), SPARTA (Zhang, et al., 2015), and docT5query (Nogueira, et al., 2019), dense retrievers such as DPR (Karpukhin, et al., 2020), ANCE (Xiong, et al., 2020), TAS-B (Hofstätter, et al., 2021), GenQ (Thakur, et al., 2021), Contriever (Izacard, et al., 2021), large language model generated embeddings such as Generalizable T5-based dense Retrievers (GTR) (Ni, et al., 2021) and cpt-text (Neelakantan, et al., 2022). Details of the compared methods are described in Appendix B.

Appendix D: Evaluation details on RELISH

Following the dataset split and evaluation settings of (Zhang, et al., 2022), we use article embeddings generated by the MedCPT article encoder to calculate the article-pair similarity and evaluate ranking quality by mean average precision (MAP) and NDCG at 5, 10, and 15.

For comparison, we also list the model performance reported in (Zhang, et al., 2022), including a random baseline, term-based retrievers such as BM25 (Robertson and Zaragoza, 2009) and PubMed Related Articles (PMRA) (Lin and Wilbur, 2007), embedding-based retrievers such as fastText (Bojanowski, et al., 2017), BioWordVec (Zhang, et al., 2019), InferSent (Conneau, et al., 2017), Sent2vec (Pagliardini, et al., 2018), BioSentVec (Chen, et al., 2019), document embedding models such as LDA (Blei, et al., 2003) and doc2vec (Le and Mikolov, 2014), and BERT-based retrievers such as BioBERT (Lee, et al., 2020), PubMedBERT (Gu, et al., 2021), SPECTER (Cohan, et al., 2020), and SciNCL (Ostendorff, et al., 2022). Details of the compared methods are described in Appendix B.

## Appendix E: Evaluation details on SciDocs

SciDocs is an evaluation framework for measuring the effectiveness of scientific paper embeddings. It includes several subtasks, such as classification of article topics, predicting user activity and citation, and also article recommendation. On SciDocs, we compare the MedCPT article encoder with various text representation models, including a random baseline, doc2vec (Le and Mikolov, 2014), FastText (Bojanowski, et al., 2017), SIF (Arora, et al., 2017), ELMo (Peters, et al., 2018), Citeomatic (Bhagavatula, et al., 2018), SGC (Wu, et al., 2019), and BERT-based models such as SciBERT (Beltagy, et al., 2019), Sent-BERT (Reimers and Gurevych, 2019), PubMedBERT (Gu, et al., 2021), SPECTER (Cohan, et al., 2020), SciNCL (Ostendorff, et al., 2022). We use the official evaluation library for SciDocs[1] and report the returned metrics for each subtask. Details of the compared methods are described in Appendix B.

Table S1 shows the evaluation results on the SciDocs benchmark. For the average performance on subtasks, the MedCPT article encoder is better than all other compared baselines except SPECTER and SciNCL. This is not surprising since (1) SciDocs includes articles in other scientific disciplines than biomedicine and (2) SPECTER and SciNCL are trained with article-article citation information and optimized for the SciDocs benchmark. However, MedCPT is still able to surpass SPECTER and SciNCL on the MeSH classification sub-task, which only contains biomedical articles. Overall, the MedCPT article encoder is comparable to SPECTER and SciNCL for scientific article representation, and better for biomedical article representation.

---

[1] https://github.com/allenai/scidocs

| SciDocs Task | Classification | | User activity prediction | | | | Citation prediction | | | | Recomm. | | Avg. |
|---|---|---|---|---|---|---|---|---|---|---|---|---|---|
| | MAG | MSH | Co-View | | Co-Read | | Cite | | Co-Cite | | | | |
| | F1 | F1 | M. | N. | M. | N. | M. | N. | M. | N. | N. | P@1 | |
| Random | 4.8 | 9.4 | 25.2 | 51.6 | 25.6 | 51.9 | 25.1 | 51.5 | 24.9 | 51.4 | 51.3 | 16.8 | 32.5 |
| Doc2vec | 66.2 | 69.2 | 67.8 | 82.9 | 64.9 | 81.6 | 65.3 | 82.2 | 67.1 | 83.4 | 51.7 | 16.9 | 66.6 |
| Fasttext-sum | 78.1 | 84.1 | 76.5 | 87.9 | 75.3 | 87.4 | 74.6 | 88.1 | 77.8 | 89.6 | 52.5 | 18.0 | 74.1 |
| SIF | 78.4 | 81.4 | 79.4 | 89.4 | 78.2 | 88.9 | 79.4 | 90.5 | 80.8 | 90.9 | <u>53.4</u> | <u>19.5</u> | 75.9 |
| ELMo | 77.0 | 75.7 | 70.3 | 84.3 | 67.4 | 82.6 | 65.8 | 82.6 | 68.5 | 83.8 | 52.5 | 18.2 | 69.0 |
| Citeomatic | 67.1 | 75.7 | 81.1 | 90.2 | 80.5 | 90.2 | 86.3 | 94.1 | 84.4 | 92.8 | 52.5 | 17.3 | 76.0 |
| SGC | 76.8 | 82.7 | 77.2 | 88.0 | 75.7 | 87.5 | <u>91.6</u> | <u>96.2</u> | 84.1 | 92.5 | 52.7 | 18.2 | 76.9 |
| SciBERT | 79.7 | 80.7 | 50.7 | 73.1 | 47.7 | 71.1 | 48.3 | 71.7 | 49.7 | 72.6 | 52.1 | 17.9 | 59.6 |
| Sent-BERT | 80.5 | 69.1 | 68.2 | 83.3 | 64.8 | 81.3 | 63.5 | 81.6 | 66.4 | 82.8 | 51.6 | 17.1 | 67.5 |
| SPECTER | **82.0** | *86.4* | <u>83.6</u> | <u>91.5</u> | <u>84.5</u> | *92.4* | *88.3* | *94.9* | *88.1* | *94.8* | **53.9** | **20.0** | <u>80.0</u> |
| SciNCL | <u>81.4</u> | <u>88.7</u> | **85.3** | **92.3** | **87.5** | **93.9** | **93.6** | **97.3** | **91.6** | **96.4** | **53.9** | *19.3* | **81.8** |
| PubMedBERT | 77.3 | 80.5 | 47.4 | 70.2 | 45.2 | 68.3 | 40.6 | 65.4 | 44.8 | 68.4 | 51.8 | 17.4 | 56.4 |
| MedCPT DEnc | *80.3* | **89.9** | *82.3* | *90.8* | *83.1* | <u>91.6</u> | 83.2 | 92.5 | *85.1* | *93.5* | *52.9* | 18.5 | *78.6* |

**Table S1.** Evaluation results of the MedCPT article encoder on the SciDocs benchmark. MeSH classification is the only biomedical task. **Bolded numbers**, <u>underlined</u>, and *italicized* numbers denote the **highest**, <u>2nd highest</u>, and *3rd highest*, respectively. All numbers are percentages. M: MAP; N: NDCG. Recomm.: recommendation. Avg.: average.

Appendix F: Evaluation details on BIOSSES and MedSTS

The evaluation is conducted under the unsupervised (zero-shot) setting, where we directly apply the MedCPT query encoder model to test set instances without any model retraining or fine-tuning. We follow the evaluation settings in (Chen, et al., 2019) and report Pearson's correlation coefficients between the model predictions and the ground truth scores. For comparison, we include SOTA methods such as BioWordVec (Chiu, et al., 2016), Universal Sentence Encoder (USE) (Cer, et al., 2018), BioSentVec trained with different corpora (Chen, et al., 2019), and BERT-based models such as PubMedBERT (Gu, et al., 2021), Clinical BERT (Alsentzer, et al., 2019), SPECTER (Cohan, et al., 2020),

SciNCL (Ostendorff, et al., 2022). Details of the compared methods are described in Appendix B.

## Appendix G: Case studies

We conduct three case studies by comparing MedCPT results with widely used web-based literature search tools, including PubMed (with Best Match ranking (Fiorini, et al., 2018)), Google Scholar[2], and Semantic Scholar[3]. Unlike the standardized biomedical IR tasks where the queries are mostly natural language sentences or questions, we choose to evaluate using short phrases (keyword combinations) that require semantic understanding in this section. We don't test on full sentences because it would be unfair for web-based search engines that are not optimized for such usage. Table S2 shows the top-3 results returned by different tools.

Query case 1 is "lead heart damage". In this query, the word "lead" most likely means the metal "lead" or the wire / cable that is used in an implanted device, but much less likely to denote the verb as in "lead to". However, "lead" is matched to "lead to" by PubMed, Google Scholar, and Semantic Scholar. MedCPT, on the other hand, matches all "lead" to the metal or a medical device. Although some titles don't explicitly contain "heart", the corresponding articles are about heart damage.

Query case 2 is "postpartum depression syndrome". In this query, the word "syndrome" is simply used to modify "postpartum depression" and does not denote any other "syndrome". The word can be neglected when retrieving relevant articles because it's not common usage. However, most web-based search engines map "syndrome" to other diseases such as "polycystic ovary syndrome" and "premenstrual syndrome" that are not part of the original information needs. MedCPT does not map the "syndrome" to other unrelated concepts, but actually tries to return more general titles that unify the query terms, i.e., "Postpartum psychiatric syndromes". MedCPT can also ignore "syndrome" as in the returned article "Postpartum Depression".

Query case 3 is "dermatologist in Germany". In this query, "Germany" should be interpreted together with "dermatologist" as the main topic. However, PubMed and Semantic Scholar mostly match "Germany" to the author's affiliation fields or the place of study. On the other hand, all three articles returned by MedCPT have exact continuous mentions of "German dermatologist(s)".

---

[2] https://scholar.google.com/
[3] https://www.semanticscholar.org/

| Query | PubMed (Best Match) | Google Scholar | Semantic Scholar | PuedCPT (ours) |
|---|---|---|---|---|
| "lead heart damage" | Brain-**Heart** Interaction: Cardiac Complications After Stroke | New treatment strategies for alcohol-induced **heart damage** | Altered Hemodynamics and End-Organ **Damage** in **Heart** Failure | Anatomical mechanisms that cause **lead** and catheter **damage** |
| | An Overview of Chemical and Biological Materials **lead to Damage** and Repair of **Heart** Tissue | An overview of chemical and biological materials **lead to damage** and repair of **heart** tissue | Hibiscus sabdariffa Linn. (Roselle) protects against nicotine-induced **heart damage** in rats | **Myocardial** changes in **lead poisoning** |
| | Role of **Cardiac** Macrophages on **Cardiac** Inflammation, Fibrosis and Tissue Repair | Early life permethrin insecticide treatment **leads to heart damage** in adult rats | Use of Computational Fluid Dynamics to Analyze Blood Flow, Hemolysis and Sublethal **Damage** to Red Blood Cells in a Bileaflet Artificial **Heart** Valve | Transvenous defibrillator **lead damage** |
| "postpartum depression syndrome" | **Depression** During Pregnancy and **Postpartum** | History of premenstrual **syndrome** and development of **postpartum depression**: a systematic review and meta-analysis | Relationship of premenstrual **syndrome** with **postpartum depression** and mother-infant bonding. | **Postpartum psychiatric syndromes** |
| | Polycystic ovary **syndrome** and **postpartum depression**: A systematic review and meta-analysis of observational studies | Associations between premenstrual **syndrome** and **postpartum depression**: a systematic literature review | Polycystic Ovary **Syndrome** and **Postpartum Depression** Symptoms: A Population-Based Cohort Study. | **Postpartum psychiatric syndromes** |
| | Relationship of premenstrual | Polycystic ovary **syndrome** and | Polycystic ovary **syndrome** and | **Postpartum Depression** |
| | **syndrome** with **postpartum depression** and mother-infant bonding | **postpartum depression**: A systematic review and meta-analysis of observational studies | **postpartum depression**: a systematic review and meta-analysis of observational studies. | |
| "dermatologist in Germany" | [Wound treatment in diabetes patients and diabetic foot ulcers] | **Dermatology in Germany** | [External scientific evaluation of the first teledermatology app without direct patient contact in **Germany** (Online Dermatologist-AppDoc)]. | [The **German** Society of **Dermatology** -- association of **German dermatologists** in Germany, Austria and Switzerland -- its position in Europe and in the world] |
| | Microbiome in healthy skin, update for **dermatologists** | To excise or not: impact of MelaFind on **German dermatologists'** decisions to biopsy atypical lesions | AI outperformed every **dermatologist** in dermoscopic melanoma diagnosis, using an optimized deep-CNN architecture with custom mini-batch logic and loss function | **German dermatologists** and their contributions to Turkish dermatology |
| | Practical management of acne for clinicians: An international consensus from the Global Alliance to Improve Outcomes in Acne | Epidemiology of contact **dermatitis**. The information network of departments of dermatology (IVDK) in **Germany** | [Why in Koenigsberg, why Samuel Jessner, why 1921? : History of the first university lectureship for sexology in **Germany**]. | [Impressions and experiences of a **German dermatologist** in America]. |

**Table S2.** Top-three retrieval article titles of MedCPT and widely used literature search engines for three case study queries. The results of PubMed, Google Scholar, and Semantic Scholar were collected on Mar 25, 2023. **Bolded** texts denote lexical matching

while **<u>bolded and underlined</u>** texts denote wrong semantic matching. Titles in "[…]" denote articles in non-English languages.

Appendix H: Scaling properties of MedCPT

In Figure S2, we study the scaling properties of MedCPT. Specifically, we evaluate the MedCPT retriever performance measured by NDCG@10 on four biomedical tasks on the BEIR benchmark (TREC-COVID, SciDocs, SciFact, NFCorpus). As shown in the figure, the performance of MedCPT increases log-linearly as the number of training logs increase, and stabilize at the end of training with 255M query-article pairs. The model needs to be trained on at least 150M query-article pairs to stabilize and consistently outperform BM25, although it should be noted that BM25 appears to be a strong baseline since many IR datasets favor BM25 due to the exposure bias in annotation. Practically, 255M query-article pairs are the most we can get from the new PubMed, and training on them already takes about 1 month of computation on a server of 8 Nvidia V100 GPUs, roughly costing ~15,000 US dollars. In conclusion, it is necessary to train on large amounts of data, but the marginal gain might decrease because the performance-training size curve follows a logarithm law.

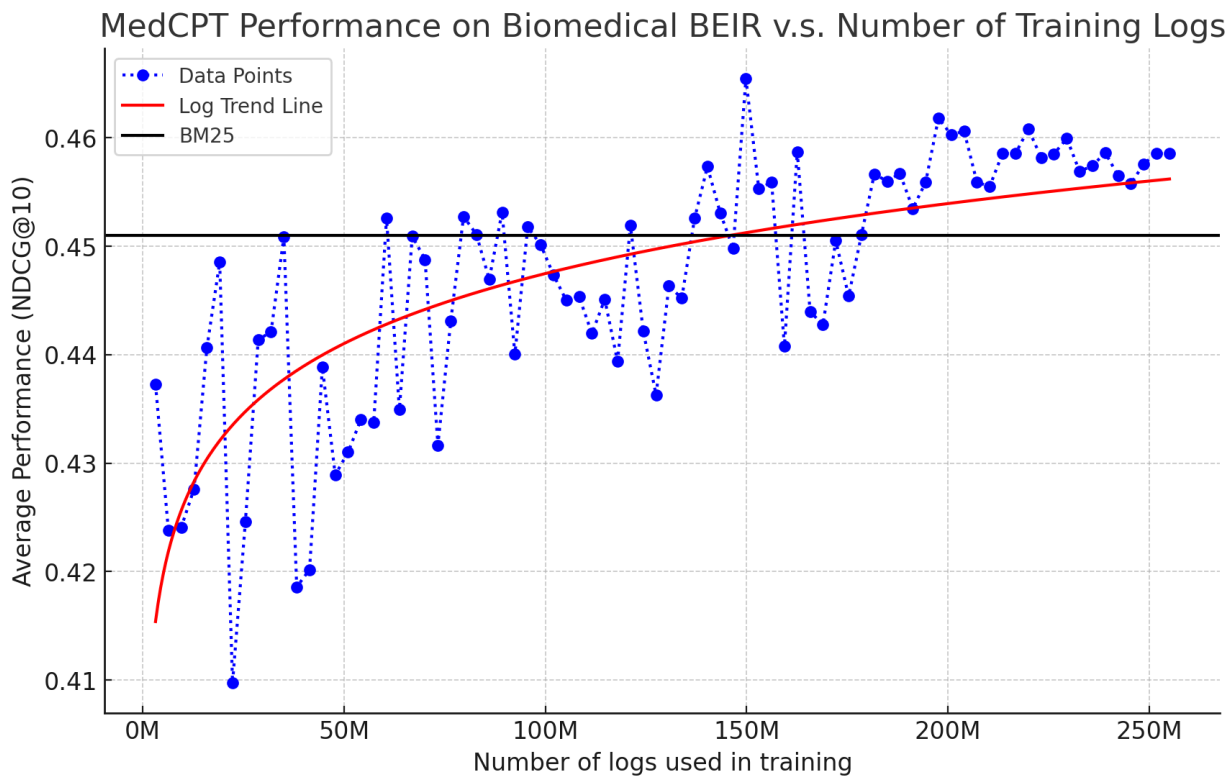

**Figure S2**. The average NDCG@10 performance on biomedical tasks in the BEIR benchmark of the MedCPT retrievers trained by different sizes of PubMed user logs. The performance increases log-linearly as the number of training logs increases.